\documentclass[a4paper,11pt]{article}
\usepackage{jcappub} 
\usepackage{lineno}
\usepackage{newtxtext,newtxmath}
\usepackage[T1]{fontenc}
\usepackage{subcaption}
 
\DeclareRobustCommand{\VAN}[3]{#2}
\let\VANthebibliography\thebibliography
\def\thebibliography{\DeclareRobustCommand{\VAN}[3]{##3}\VANthebibliography}
 
\newcommand{\be}{\begin{eqnarray}}
\newcommand{\ee}{\end{eqnarray}}
\newcommand{\beq}{\begin{equation}}
\newcommand{\eeq}{\end{equation}}
\newcommand{\exclude}[1]{}

\usepackage{xcolor}

\newcounter{LVWcounter}

\newcounter{FMcounter}

\newcounter{XLcounter}

\def\rmd{\mathrm{d}}

\title{The Glow of Axion Quark Nugget Dark Matter: (IV) CMB Spectral and Anisotropy Signatures}

\author[a]{Fereshteh Majidi,}
\author[a]{Xunyu Liang,}
\author[a, b]{Michael Sekatchev,}
\author[a]{Ludovic Van Waerbeke,}
\author[a]{Ariel Zhitnitsky}
\affiliation[a]{Department of Physics and Astronomy,\\
University of British Columbia, \\
V6T 1Z1 Vancouver, BC, Canada}
\affiliation[b]{Department of Nuclear Engineering, \\
University of California, Berkeley, \\
94720 Berkeley, CA, USA}

\emailAdd{fereshtehmajidi@phas.ubc.ca}

\abstract{
Axion quark nuggets (AQNs) are macroscopic dark-matter candidates, with masses of the order of a few grams to a kilogram and sub-micron radius, thought to form at the Quantum Chromo Dynamic era through axion-induced charge separation. This framework naturally links the dark and visible matter abundances ($\Omega_{\rm DM}\sim \Omega_{\rm visible}$) and provides a mechanism for generating the baryon–antibaryon asymmetry where dark matter is composed of matter AQNs and antimatter AQNs. Although behaving as cold dark matter on cosmological scales, baryons annihilate with antimatter AQNs, producing ionizing high-energy photons.
The resulting energy injection may imprint spectral distortions on the cosmic microwave background (CMB) and modify the reionization history.
Using a modified version of the \texttt{CLASS} Boltzmann code we compute the impact of this energy injection on the $\mu$ and $y$ spectral distortion parameters as well as on the optical depth. We find that (1) the CMB anisotropies remain essentially unaffected by baryon annihilation, and (2) the associated spectral distortion signatures lie within the sensitivity reach of proposed CMB spectral distortion missions. Finally, we discuss the similarities and differences between the AQN scenario and annihilating or decaying dark matter models.
}

\keywords{Dark matter theory, Dark matter signatures, Cosmic microwave background theory and observations} 
\begin{document}
\maketitle
\flushbottom

\section{Introduction}

Since the groundbreaking observations of the Cosmic Background Explorer (COBE) in the early 1990s \cite{Fixsen:1996nj}, research on the Cosmic Microwave Background (CMB) has primarily focused on temperature and polarization anisotropies. However, distortions of the CMB energy spectrum, relative to a perfect blackbody, can provide a powerful and complementary probe of the early Universe, capable of tracing energy-release processes across a broad range of cosmic epochs \cite{Chluba:2011hw, 2013IJMPD..2230014S}.

Such spectral distortions can arise at redshift $z<2\times 10^6$, caused by processes that can drive matter and radiation out of equilibrium, while the time available before recombination is insufficient for full thermalization. For instance, energy injection by dark matter decay or annihilation, primordial black hole evaporation, and dissipation of acoustic waves lead to distinct spectral signatures. Aboard COBE, FIRAS remains the only instrument to have measured the absolute frequency spectrum of the CMB with high precision, reaching a relative accuracy of $\sim10^{-5}$ \cite{Mather:1993ij}. However, this precision remains insufficient to detect the subtle spectral distortions expected from incomplete thermalization; as of today, the CMB frequency spectrum remains fully consistent with a pure blackbody. With proposed missions like PRISM \cite{2014JCAP...02..006A}, ESA's Voyage2050 \cite{2021ExA....51.1515C}, PIXIE \cite{2011JCAP...07..025K,2020JCAP...05..041K,Kogut:2024vbi}, FOSSIL \cite{FOSSIL2022}, the field is poised for a major leap forward  \cite{2024EPJWC.29300012C}, with a projected improvement of four to five orders of magnitude over FIRAS's spectral sensitivity. It will be possible to test annihilating or decaying dark matter \cite{2013MNRAS.436.2232C,2015PhRvL.115g1304A} and more exotic models \cite{Chluba:2011hw,2024JCAP...07..019L,2019PhRvD..99l3510A,2025PhRvD.111j3516G}.

In this work, we investigate the CMB spectral distortions produced by macroscopic dark matter composed of objects far heavier than the Planck mass, which emit photons from their surfaces and inject energy into the baryon–photon plasma prior to recombination. The possibility that macroscopic dark matter could contribute to spectral distortions was first suggested in \cite{Kumar:2018rlf}. Here, however, we focus on a specific and physically motivated model: the Axion Quark Nugget (AQN) scenario \cite{Zhitnitsky:2002qa}. AQNs are formed during the quark–hadron phase transition in the epoch of matter–antimatter annihilation. Their formation arises naturally within Quantum Chromodynamics (QCD) in the presence of a QCD axion field, which is motivated by the strong CP problem. The asymmetric coupling of the axion to matter and antimatter (due to the CP-odd nature of the axion field) induces an asymmetric charge-separation mechanism that produces quark nuggets and antiquark nuggets. The antiquark component of the dark matter can annihilate with ordinary matter, depositing energy into the surrounding medium. Previous work has shown that this energy injection produces a faint but potentially detectable “dark matter glow’’ in large-scale structures \citep{Majidi:2024mty}, galaxy clusters \citep{Sommer:2024iqp}, and may help explain the puzzling far-ultraviolet emission observed in the Milky Way \citep{Sekatchev:2025ixu}. In this paper, we extend the analysis to the pre-recombination era to investigate the impact of this energy injection on CMB anisotropies and the CMB frequency spectrum.

To capture the physics of departures from equilibrium, we use the \texttt{CLASS} code ({\it Cosmic Linear Anisotropy Solving System}), a widely used Boltzmann solver that computes the evolution of cosmological perturbations for large-scale structure and the CMB, as well as the CMB frequency spectrum \cite{Blas:2011rf, Lucca:2019rxf}. We modify \texttt{CLASS} to incorporate the energy injection associated with the AQN dark matter model. This work discusses the physics of AQNs, the implementation of these effects in \texttt{CLASS}, and the resulting impact on CMB observables.

The paper is organized as follows. In Section \ref{sect:AQN}, we present a short overview of the AQN model and its energy injection mechanism. Section \ref{sec:CMB signatures of energy injection} discusses the CMB signatures (spectral distortions and anisotropies) of AQNs and conventional decaying/annihilating dark matter. Section \ref{sec:Results} presents our results, and Section \ref{sec:Discussion and Conclusion} summarizes our conclusions and outlines directions for future work.

\section{The Axion Quark Nuggets Model}\label{sect:AQN}

The AQN model~\cite{Zhitnitsky:2002qa} (see also a brief overview in \cite{Zhitnitsky:2021iwg}  and lectures \cite{VanWaerbeke:2026sxs}), was proposed to explain why the mass densities of dark matter (DM) and visible matter are of similar magnitude. Drawing parallels with Witten’s quark nugget framework, AQNs also qualify as a form of cold dark matter (CDM) because they were non-relativistic at the formation time and effectively remain collisionless today. The key requirement for dark matter to be considered collisionless is that the self-interaction rate is low enough so that thermalization is not possible and the mean free path is much larger than the size of the largest observed structures. In practice, this requires that the ratio of the self-interaction cross section $\sigma_{\rm DM-DM}$ to its mass $M_{\rm DM}$ must be sufficiently small: 

\be
\label{sigma/m}
\frac{\sigma_{\rm DM-DM}}{M_{\rm DM}} \lesssim 1 \;{\rm cm^2g^{-1}}.
\ee

For Weakly Interacting Massive Particles (WIMPs), this ratio is $\sim 10^{-14}\;{\rm cm^2g^{-1}}$ \cite{2018PhR...730....1T}, which easily satisfies condition (\ref{sigma/m}) due to their extremely small interaction cross-section for a typical mass $m_{\rm WIMP}\in( 10^2-10^3) ~\rm GeV$. In contrast, for AQNs, the ratio is $\sim 10^{-10}\;{\rm cm^2g^{-1}}$, which also satisfies condition (\ref{sigma/m}). In this case, however, the small value arises primarily from their extremely large mass rather than from a suppressed cross-section.
As we discuss below, the interaction cross-section of AQNs is large, since they are composed of ordinary quarks, antiquarks, and gluons from Standard Model particle physics. Other conditions for collisionless DM require that the DM-baryon and DM-photon interactions are also small, generally much smaller than (\ref{sigma/m}). The corresponding cross-sections are constrained by the damping of acoustic oscillations in the CMB anisotropies, $\sigma_{\rm DM-bar}/M_{\rm DM}\lesssim  0.003\;{\rm cm^2g^{-1}}$ \cite{Dvorkin:2013cea} and $\sigma_{\rm DM-\gamma}/M_{\rm DM}\lesssim  4.5\times10^{-7}\;{\rm cm^2g^{-1}}$ \cite{2014JCAP...04..026W}. One objective of this work is to quantify the impact on the CMB anisotropies and frequency spectrum through a rigorous calculation of the AQN–baryon and AQN–photon interactions using a Boltzmann code.

We refer the reader to the original AQN papers \cite{Liang:2016tqc,Ge:2017ttc,Ge:2017idw,Ge:2019voa}, which address in detail the specific questions related to nugget formation, the generation of the baryon asymmetry, the survival of the nuggets throughout the evolution of the early Universe, and the model’s consistency with current observational constraints. In this framework, a fraction of the AQNs are composed of antimatter, ensuring that the total baryon charge of the Universe remains zero. This naturally leads to dark matter and visible matter density parameters of the same order of magnitude, $\Omega_{\rm DM}\sim \Omega_{\rm visible}$. The absolute stability of AQNs in vacuum is built into the model: the energy per baryon in quark matter is smaller than in hadronic matter, preventing decay into ordinary baryons and guaranteeing their persistence to the present day.

The presence of antimatter nuggets makes AQNs strongly interacting objects (i.e., with a large cross-section), as they annihilate upon contact with surrounding matter. This property is particularly relevant for the present work, since interactions between antimatter AQNs and ordinary matter can produce a distinctive AQN signature in the CMB frequency spectrum. Long before recombination, such annihilation events would inject energy into the primordial plasma, potentially altering the thermal history of the Universe and leaving observable imprints on the CMB. In particular, the annihilation of antimatter AQNs with the baryonic medium releases electromagnetic radiation, including X-rays, which could distort the CMB spectrum through $\mu$ and $y$ type distortions. The efficiency and magnitude of this energy injection depend on the AQN number density, mass, and effective cross-section, the latter of which can be significantly enhanced in ionized environments through Coulomb interactions.

For the purpose of this work, we focus on the high-energy regime relevant to CMB photons, where the \texttt{CLASS} code provides a robust framework for modelling energy injection. Although \texttt{CLASS} treats the energy injection as frequency-independent, this approximation is well suited for studying the incomplete thermalization of the CMB from high-energy photon injection, particularly the generation of $\mu$-type distortions at redshifts $z \gtrsim 5 \times 10^4$ and $y$-type distortions at later times when Compton scattering becomes inefficient. Incorporating the AQN dark matter model into \texttt{CLASS} enables us to explore how high-energy electromagnetic signatures from AQN annihilation events alter the thermal and ionization history of the Universe, and to make quantitative predictions about their impact on the CMB power spectra. By accurately modeling the heating history from AQNs in this high-energy context, we aim to assess their potential role as a dark matter candidate and their observational consequences in current and future CMB experiments.

\subsection{Characteristics of AQNs}
\label{subsec:Characteristics of AQNs}
While both matter and antimatter AQNs exist, only the antimatter AQNs produce appreciable energy injection through annihilation with ordinary matter. For simplicity, we refer to ``antimatter AQNs'' as ``AQNs'' in the remainder of this paper. We adopt the same convention as in \cite{Majidi:2024mty,Sommer:2024iqp,Sekatchev:2025ixu} and define the number density of AQNs as: 
\begin{equation} 
\begin{aligned}
n_{\rm AQN}
\equiv\frac{2}{3}\times\frac{3}{5}\times\frac{\rho_{\rm DM}}{m_{\rm AQN}}
&
=9.02\times10^{-33}(1+z)^3{\rm\,cm^{-3}}
\left(\frac{100{\rm\,g}}{m_{\rm AQN}}\right)\,,
\end{aligned}
\end{equation} 
where the factor of $\frac{2}{3}$ accounts for the electromagnetic fraction of the AQN emission, the factor of $\frac{3}{5}$ represents the antimatter sector of the AQNs, $\rho_{\rm DM}$ is the DM mass density, and $c$ is the speed of light in vacuum. For simplicity, we assume that AQNs constitute the dominant component of dark matter and that all AQNs have the same mass, $m_{\rm AQN}$. 

Assuming AQNs have nuclear density and a mass in the range of 10-1000\,g (see, e.g.\cite{PandaX:2022xqx,LZ:2022lsv,XENON:2023cxc, Majidi:2024mty,Sommer:2024iqp,Sekatchev:2025ixu}), the results in Section \ref{sec:Results} show that variations of $m_{\rm AQN}$ within the allowed mass range have a negligible impact on the resulting constraints. Motivated by this weak dependence, we fix $m_{\rm AQN}$ to a representative value throughout this work. However, we can relate the geometrical radius $R$ of an AQN with its mass $m_{\rm\,AQN}$:
\begin{equation}
\begin{aligned}
R 
&
\approx4.09\times10^{-5}
{\rm\,cm}\left(\frac{m_{\rm AQN}}{100\rm\,g}\right)^{1/3}.
\end{aligned}
\end{equation}

The geometrical radius $R$ sets the lower bound on the AQN cross section in interactions with ordinary matter. The geometrical cross section, $\pi R^2$, corresponds to the interaction cross section with neutral particles (e.g., atomic hydrogen). The interaction cross section with baryons can be substantially larger when the particles are charged (e.g., protons H$^+$). This is because AQNs are charged in a plasma, and they attract charged particles via Coulomb interaction. In the early Universe, the charge distribution in the vicinity of an AQN was estimated in \cite{Flambaum:2018ohm}:
\begin{equation}
n(r)
\approx\frac{m_ek_{\rm B}^2T^2}{\pi\alpha\hbar^3 c}
\left(\frac{R}{r}\right)^6\,,
\end{equation}
where $k_{\rm B}$ is the Boltzmann constant, $\alpha$ is the fine-structure constant, $\hbar$ is the reduced Planck constant, $c$ is the speed of light, $m_e$ is the electron mass, $T$ is temperature of the plasma, $r$ is the distance from the centre of the AQN. For a given effective capture radius $R_{\rm eff}$, the effective charge ${\cal Q}_{\rm eff}$ within the capture radius is:
\begin{equation}
\label{eq:cal Q_eff R_eff}
{\cal Q}_{\rm eff}(R_{\rm eff})
\approx\int_{R_{\rm eff}}^\infty 4\pi r^2n(r)\rmd r
=\frac{4 k_{\rm B}^2m_e R^3 T^2}{3\alpha\hbar^3c}
\left(\frac{R}{R_{\rm eff}}\right)^3\,.
\end{equation}

To first order, the effective capture radius, $R_{\rm eff}$, is defined as the distance at which the Coulomb potential equals the thermal kinetic energy:
\begin{equation}
\label{eq:alpha hbar c Q_eff R_eff}
\frac{\alpha\hbar c {\cal Q}_{\rm eff}}{R_{\rm eff}}
\approx k_{\rm B} T\,.
\end{equation}
Solving for eqs. \eqref{eq:cal Q_eff R_eff} and \eqref{eq:alpha hbar c Q_eff R_eff}, we obtain the explicit formula of $R_{\rm eff}$:
\begin{equation}
\label{eq:R_eff R 2}
\begin{aligned}
\left(\frac{R_{\rm eff}}{R}\right)^2
\approx\frac{R}{\hbar}\sqrt{\frac{4m_ek_{\rm B}T}{3}}
&
=936\left(\frac{k_{\rm B}T}{0.3{\rm\,eV}}\right)^{1/2}
\left(\frac{m_{\rm AQN}}{100{\rm\,g}}\right)^{1/3}\,.
\end{aligned}
\end{equation}

Even at a relatively late time in the early Universe, such as the recombination epoch ($k_{\rm B}T \approx 0.3\,{\rm eV}$), the effective cross section remains two to three orders of magnitude larger than the geometrical cross section. This result will play a central role in the present work, as will be discussed below.

The derivation of $R_{\rm eff}$ in this subsection is valid in a plasma as long as the screening effect remains negligible. Otherwise, the effective charge in eq. \eqref{eq:alpha hbar c Q_eff R_eff} will be suppressed by Debye screening, i.e. $Q_{\rm eff}\rightarrow Q_{\rm eff}e^{-R_{\rm eff}/\lambda_{\rm D}}$, where $\lambda_{\rm D}$ is the Debye length. As shown in appendix \ref{app:A check of screening effect}, this assumption ($R_{\rm eff}\ll\lambda_{\rm D}$) remains valid within the redshift range of interest for this work ($10^3\lesssim z\lesssim10^6$). At even higher redshifts $z\gtrsim10^8$ where the screening effect becomes critical, the effective capture radius is still much greater than the geometrical size $R$, but should be defined differently \cite{Flambaum:2018ohm,Ge:2019voa}.

\subsection{AQN energy injection in the early Universe}
\label{subsec:AQN energy injection in the early Universe}
The annihilation of antimatter within the AQNs, resulting from the collision with ordinary matter, produces a wide range of electromagnetic radiation from radio to $\gamma$-rays \cite{Forbes:2006ba,Forbes:2008uf,Lawson:2013bya}. The thermal history of the early Universe, from approximately a few months after the Big Bang to the present, is modified by the injection of external photons and their interactions with ambient particles. For AQN energy injection, we consider the X-ray band as the dominant emission channel.
This type of emission is assumed to be less than 50\% of the total annihilation energy \cite{Forbes:2006ba}. We will denote $g$ as the fraction of injection energy in the total annihilation energy, and assume $0.1\lesssim g\lesssim0.5$ \cite{Forbes:2006ba}.

It is straightforward to derive the energy injected in the environment from the annihilation with a single proton ($\rm H^+$). Using eq. \eqref{eq:R_eff R 2}, the energy injection rate by AQNs per unit volume, $\dot{Q}_{\rm approx}$ $\rm [erg\,s^{-1}\,cm^{-3}]$, is given by:
\begin{equation}
\label{eq:dot Q_apprx}
\dot{Q}_{\rm approx}
\equiv X_e\left(\frac{R_{\rm eff}}{R}\right)^2\dot{Q}_0\,,
\end{equation}
where $E_{\rm ann}\approx2{\rm\,GeV}$ is the annihilation energy of a $p\bar{p}$ pair, the electron fraction $X_e\equiv n_{\rm e}/n_{\rm H,tot}$ is the ratio of the number density of free electron $n_e$ to the total number density of hydrogen $n_{\rm H,tot}$ (neutral and ionized),
$n_{\rm b}$ is the baryon number density, $\rm \Delta v$ is the relative speed between the dark matter and the baryon, and $Q_0$ is defined to ``normalize'' the energy injection rate:
\begin{equation}
\label{eq:Q_0}
\dot{Q}_0
\equiv g \pi R^2E_{\rm ann}n_{\rm AQN}n_{\rm b}{{\rm \Delta v}}\,.
\end{equation}

It is important to emphasize that the energy-injection rate (\ref{eq:dot Q_apprx}) is derived from the specific physics of AQNs, whereas the prescription in \cite{Kumar:2018rlf} is intentionally model-agnostic and does not describe any particular macroscopic–DM scenario. A further key distinction is that, for AQNs, the energy injection (\ref{eq:Q_0}) scales with the product of the baryonic and DM number densities, i.e.\ $\propto (n_{\rm AQN}\cdot n_{\rm b})$. By contrast, for particle dark matter the injection typically scales as $\propto n_{\rm DM}$ (decays) or $\propto n_{\rm DM}^2$ (annihilations) \footnote{For the macros considered in \cite{Kumar:2018rlf}, the energy source is assumed to arise from the DM component alone, effectively behaving like decaying DM, i.e.\ $\propto n_{\rm DM}$.}.

For a more realistic approach, the early Universe does not only consist of $\rm H^+$, but we also have to consider all other ionized species. For example, neutral hydrogen H becomes the dominant species shortly after recombination. Also, helium-4 ($\rm ^4He$, $\rm ^4He^+$, and $\rm ^4He^{++}$) contributes about 25\% of the baryon sector, while other elements excluding hydrogen and helium-4 comprise less than 1\%. We will adopt a generalized formula of the energy injection rate per unit volume, $\dot{Q}$ $\rm [erg\,s^{-1}\,cm^{-3}]$, that must take into account the diversity of ions:

\begin{equation}
\label{eq:dot Q}
\dot{Q}
=\sum_i\dot{Q}_i
\equiv \left(\frac{\sigma}{\pi R^2}\right)\dot{Q}_0\,,\qquad
\dot{Q}_i
=f_ig\sigma_i E_{{\rm ann},i} n_{\rm AQN}n_i{{\rm \Delta v}_i}\,,
\end{equation}
where $i$ indexes the species $\mathrm{H}$, $\rm H^+$, $\rm ^4He$, $\rm ^4He^+$, and $\rm ^4He^{++}$, as appropriate for the cosmological epoch. The quantity $\sigma$ denotes the total cross section, and the weighting includes the annihilation energy per collision $E_{\rm ann,i}$, the annihilation probability $f_i$, the number density $n_i$, and each species’ relative speed with respect to the dark matter, ${\rm\Delta v}_i$. The additional factor $f_i$ accounts for the reduced annihilation probability of the neutral species $\mathrm{H}$ and $\rm ^4He$, for which $f_{i,{\rm neu}}\approx0.1$ per collision \cite{Forbes:2006ba}. For the ionized species ($\rm H^+$, $\rm ^4He^+$, and $\rm ^4He^{++}$), we take $f_{i,{\rm ion}}\approx1$ because charged particles are captured by the Coulomb potential and will orbit within the capture radius until annihilation occurs.

As derived in appendix \ref{app:Weighted cross section sigma}, the weighted cross section in the early Universe is then given by:
\begin{equation}
\label{eq:sigma pi R 2}
\frac{\sigma}{\pi R^2} \approx (1 - Y_p) X_e \left( \frac{R_{\rm eff}}{R} \right)^2 + 0.1 \left [ (1 - Y_p)(1 - X_e) + \frac{1}{2}Y_p \right],
\end{equation}
where $X_e$ is the ionization fraction and $Y_p \approx 0.245$ is the primordial helium mass fraction. More specifically, $X_e$ and $Y_p$ are defined as:
\begin{equation}
X_e\equiv\frac{n_e}{n_{\rm H,tot}}\,,\qquad
Y_p\equiv\frac{4n_{\rm He,tot}}{4n_{\rm He,tot}+n_{\rm H,tot}}
=\frac{4n_{\rm He,tot}}{n_{\rm b}}\,,
\end{equation}
where $n_e$ is the free electron number density, $n_{\rm H,tot}$ (or $n_{\rm He,tot}$) is the number density of all neutral and ionized H (or respectively, $\rm ^4He$).

We use both energy-injection rates, $\dot{Q}_{\rm approx}$ and $\dot{Q}$. In section \ref{sec:AQN-induced signatures}, we adopt the approximate expression \eqref{eq:dot Q_apprx} for $\dot{Q}_{\rm approx}$ to obtain analytic estimates and highlight key features of the AQN signatures. In section \ref{sec:Results}, we employ the full expression for $\dot{Q}$ given in \eqref{eq:dot Q} in our numerical estimates. We then compare the analytically predicted features with the numerical results.

\section{CMB signatures of energy injection}
\label{sec:CMB signatures of energy injection}

\subsection{General theory of energy injection}

In this section, we outline the mechanisms linking energy injection to spectral distortions. Detailed derivations can be found in \cite{Zeldovich:1969, Sunyaev:1970, Zeldovich:1971, Chluba:2011hw}. Because we consider only the injection of high-energy photons, the approximation in which the injected energy is treated independently of the photon frequency is adequate; this is the approach implemented in \texttt{CLASS}. For low-energy photons, however, atomic-level interactions must be taken into account -- e.g., using \texttt{CosmoTherm} -- which we leave to future work. Low-energy photons are unlikely to influence pre-recombination physics, as most of the matter is ionized, although they become important during the dark ages and cosmic dawn epochs, which lie outside the scope of this study. Our treatment is therefore analogous to that used for energy injection from annihilating or decaying dark matter, where only the total injected energy is relevant, and the photon energy distribution is not explicitly followed. For this reason, in the remainder of the paper, we often compare the AQN dark-matter signatures to those of annihilating or decaying dark-matter models.

\subsubsection{Spectral distortions}
In the standard cosmological model, the CMB follows an almost perfect blackbody spectrum, but small spectral distortions can arise from processes that inject or redistribute energy in the photon-baryon fluid. The standard approach, first developed by \cite{Zeldovich:1969, Sunyaev:1970, Zeldovich:1971}, consists of a parametrization of the deviations from a pure blackbody using two types of spectral distortions: the $\mu$-type and $y$-type. Note that spectral distortions can be effectively developed only for $z<2\times 10^6$, because any energy injection at higher redshift will have time to fully thermalize, resulting in a pure blackbody spectrum with a different photon temperature which cannot be observed. The $\mu$-distortion is generated at early times ($z \gtrsim 5 \times 10^4$), when Compton scattering efficiently redistributes energy but photon number-changing processes become inefficient, resulting in a nonzero chemical potential in the photon distribution. At later times ($z \lesssim 10^4$), when Compton scattering itself becomes inefficient, energy injection leads to a $y$-distortion, which changes the shape of the CMB spectrum without altering the photon number. Precise measurements of CMB spectral distortions offer a unique window into nonstandard energy injection mechanisms, such as decaying or annihilating particles, primordial black holes, or exotic dark matter models like axion quark nuggets, providing critical constraints on physics beyond the standard model.

The magnitude of the resulting spectral distortions is entirely governed by the thermal history of the Universe and can be parameterized through the heating rate, $\Dot{Q}$ $\rm [erg\,s^{-1}\,cm^{-3}]$.
The quantities $\frac{\rmd(Q/\rho_\gamma)}{\rmd z'}$ (the energy release relative to CMB blackbody) and $\mathcal{J}(z')$ (the visibility function) play a fundamental role in understanding how thermal processes shape the CMB. They are essential for identifying the epochs when various types of spectral distortions become most prominent and for characterizing not only standard thermal effects but also additional energy injection sources, such as AQNs. By analyzing these functions, we gain valuable insights into the signatures left by early-Universe physics on the CMB, providing important probes of non-standard cosmological scenarios.

In these calculations, the number density of photons plays a critical role, as it is affected by radiative processes that alter the chemical potentials $\mu$, $\nu$, and the temperature, alongside exotic heating mechanisms that can vary depending on the specific dark matter model under consideration \cite{Lucca:2019rxf}. The total intensity distortion can be expressed as 
\begin{equation}
\label{eq:Delta I_tot nu z}
    \Delta I_{\mathrm{tot}}(\nu, z) = \int_{z}^{\infty} \rmd z'\, G_{\mathrm{th}}(\nu, z')\, \frac{\rmd Q(z')/\rmd z'}{\rho_\gamma(z')},
\end{equation}
where the thermal Green's function $G_{\mathrm{th}}(\nu, z')$ is given by 

\begin{equation}
    G_{\mathrm{th}}(\nu, z') = \mathcal{G}(\nu)\, \mathcal{J}_g(z') + \mathcal{Y}(\nu)\, \mathcal{J}_y(z') + \mathcal{M}(\nu)\, \mathcal{J}_\mu(z') + R(\nu, z').
\end{equation} 
where $\mathcal{G}(\nu)$ denotes the spectral shape corresponding to a pure blackbody temperature shift, with $\mathcal{J}_g(z')$ representing its redshift-dependent amplitude. The function $\mathcal{Y}(\nu)$ describes the characteristic Compton $y$-distortion spectrum, arising from inverse Compton scattering between CMB photons and hot electrons, and $\mathcal{J}_y(z')$ is its associated amplitude. The term $\mathcal{M}(\nu)$ corresponds to the $\mu$-distortion spectrum, which captures the effect of energy release at early times ($z \gtrsim 10^5$) resulting in a Bose–Einstein–like distribution with a chemical potential $\mu$; its amplitude is denoted by $\mathcal{J}_\mu(z')$. Finally, $R(\nu, z')$ accounts for the residual distortion component not captured by the standard basis $\mathcal{G}, \mathcal{Y}, \mathcal{M}$ (see equation (3.7) in \cite{Lucca:2019rxf}).
This residual encodes information about energy-release histories that deviate from the standard templates and can be essential for identifying exotic processes such as dark matter decay, annihilation, or AQNs. As noted earlier, the energy injection in \texttt{CLASS} is implemented without frequency dependence, which limits its applicability primarily to ionizing (high-energy) photons.  

\paragraph{Chemical potential $\mu$-distortion}
One important example of a spectral distortion is the chemical potential, or $\mu$, distortion, which arises when energy is injected into the photon-baryon fluid at redshifts $z \gtrsim 5 \times 10^4$. During this era, Compton scattering efficiently redistributes energy among photons, but photon number-changing processes, such as double Compton scattering and Bremsstrahlung, become inefficient. As a result, the photon distribution deviates from a pure blackbody and develops a nonzero chemical potential.
 
The $\mu$-distortion arises from thermal processes acting on the photon distribution, which can be described using the Boltzmann equation. Assuming the photon occupation number takes the form $n \approx \frac{1}{e^x - 1} - \mu(z, x) \frac{G(x)}{x}$, the evolution of the distortion is governed by
\begin{equation}
0 \approx \frac{\theta_e}{x^2} \frac{\partial}{\partial x} \left( x^4 \left[ \frac{\partial n}{\partial x} + \frac{T_\gamma}{T_e} n (1+n) \right] \right) + \frac{K}{x^3} \left( 1 - n(e^x - 1) \right),
\end{equation}
where $\theta_e$ is the dimensionless electron temperature, $x = h\nu / kT_\gamma$ is the dimensionless frequency, and $K$ characterizes photon production processes such as double Compton scattering and Bremsstrahlung. This leads to a chemical potential distortion of the form
\begin{equation}
\mu(z, x) = \mu_0 \cdot e^{-x_c(z)/x},
\end{equation}
where $\mu_0$ is the initial amplitude and $x_c(z)$ is the critical frequency marking the transition between different thermal regimes. The amplitude is approximately
\begin{equation}
\mu_0 \approx 1.401 \frac{\Delta \rho_\gamma}{\rho_\gamma} \approx 1.401 \int \frac{\rmd(Q/\rho_\gamma)}{\rmd z'} \mathcal{J}(z') \, \rmd z',
\label{eq:mu0}
\end{equation}
where $\frac{\rmd(Q/\rho_\gamma)}{\rmd z'}$ describes the energy release relative to the CMB blackbody energy density, and $\mathcal{J}(z')$ is the visibility function that quantifies the fraction of injected energy at redshift $z'$ that survives to contribute to today’s observable distortion~\cite{Chluba2014,Lucca:2019rxf}. Together, these quantities provide a framework for connecting early-Universe energy injection mechanisms, such as those from axion quark nuggets, to measurable signatures in the CMB spectrum.

\paragraph{Compton $y$-distortion}
Another important type of spectral distortion is the Compton \( y \)-distortion, which arises when energy is injected at redshifts below \( z \lesssim 10^4 \), during an era when Compton scattering is still happening but is no longer efficient enough to establish full equilibrium before recombination. In this regime, while the photon distribution is redistributed by scattering, it does not reach a Bose-Einstein spectrum with a chemical potential, as in the case of the \(\mu\)-distortion. Instead, the interaction leads to a characteristic \( y \)-distortion, where the total energy in the photon gas increases, the photon number remains conserved, and the spectrum acquires a tilt (fewer low frequency photons and more high frequency photons with a pivot frequency at $\nu=217\;{\rm GHz}$).

\subsubsection{Anisotropies}

Energy injection during and after recombination also affects the CMB anisotropies \cite{Adams:1998nr,Chen:2003gz,Padmanabhan:2005es}. The main effect is that the additional heating raises the ionization fraction $X_e$ and enhances the optical depth; consequently, the amplitude of the anisotropy spectrum can change. Energetic photons above the Lyman-$\alpha$ threshold interact efficiently with the gas. In the study \cite{Padmanabhan:2005es}, the most efficient ionization occurs for photons with energy $\lesssim50{\rm\,keV}$ at redshift $z\sim1000$. Above $50{\rm\,keV}$, photon interactions are dominated by processes such as Compton scattering, pair production, and photon–photon scattering. Photons in the energy range $\sim10^8$--$10^{10}{\rm\,eV}$ may even freeze out, since the photon cooling time exceeds the Hubble time, allowing them to decouple from the gas. Effective ionization through external photon injection broadens the surface of last scattering and modifies the temperature and the polarization anisotropies of the CMB  \cite{Chen:2003gz,Padmanabhan:2005es}.

Exotic injection can come from either annihilating DM (e.g., \cite{Padmanabhan:2005es,Slatyer:2009yq,Kanzaki:2009hf,Galli:2009zc,Hisano:2011dc,Hutsi:2011vx,Galli:2011rz,Slatyer:2012yq,Galli:2013dna,Madhavacheril:2013cna,Slatyer:2015kla,Slatyer:2015jla,Finkbeiner:2011dx,Cang:2020exa}) or decaying DM (e.g., \cite{Adams:1998nr,Chen:2003gz,Zhang:2007zzh,Finkbeiner:2011dx,Slatyer:2016qyl,Poulin:2016anj,Acharya:2019uba,Cang:2020exa,Capozzi:2023xie,Liu:2023nct,Xu:2024vdn}). In both cases, the DM particles inject energetic photons and alter the ionization and thermal history of the Universe. In terms of energy injection rate per unit volume ($\dot{Q}_{\rm ann}$ and $\dot{Q}_{\rm dec}$), the two models have distinct redshift dependence of energy:
\begin{subequations}
\label{eqs:dot Q_ann etc}
\begin{equation}
\label{eq:dot Q_ann}
\dot{Q}_{\rm ann}
=p_{\rm ann}\rho_{\rm crit,0}^2c^4\Omega_{\rm DM,0}^2(1+z)^6\,,\qquad
\textrm{(DM annihilation)}
\end{equation}
\begin{equation}
\dot{Q}_{\rm dec}
=p_{\rm dec}\rho_{\rm crit,0}c^2\Omega_{\rm DM,0}(1+z)^3\,,\qquad
\textrm{(DM decay)}
\end{equation}
\end{subequations}
where $p_{\rm ann}$ and $p_{\rm dec}$ are referred as the ``energy deposition yield'' in \cite{Finkbeiner:2011dx} that describe all information about the source of energy injection and the efficiency with which that energy ionizes the gas; $\rho_{\rm crit,0}$ is the critical mass density of the present day; $\Omega_{\rm DM,0}$ is the normalized DM mass density of the present day; $c$ is the speed of light.

At higher redshifts $z\gtrsim1000$, the energy injection has a negligible impact on the CMB anisotropies when the Universe is fully ionized. At lower redshifts, the energy deposition may become insignificant if the DM injection rate rises steeply with redshift. DM annihilation injects energy much more efficiently at early times than does DM decay. A study in \cite{Finkbeiner:2011dx} suggests that the impact of DM annihilation peaks around $z\approx600$, whereas DM decay continues to influence the CMB throughout the post-recombination epoch.

The constraint on $p_{\rm ann}$ from the Planck \cite{Planck:2018vyg} is given by:
\begin{equation}
\label{eq:p_ann}
p_{\rm ann}
\approx\left. f(z)\frac{\langle\sigma v\rangle_{\rm ann}}{m_\chi c^2}\right|_{z=600}
\lesssim3.2\times10^{-28}{\rm\,cm^3s^{-1}GeV^{-1}}\,,\quad
\textrm{(95\% C.L.)}
\end{equation}
where $f(z)$ is the ``efficiency parameter" of the model, $\langle\sigma v\rangle_{\rm ann}$ is the thermally averaged annihilation cross section times the velocity, and $m_\chi$ is the mass of the DM particle. The current CMB probes are sensitive to a redshift range of $z\approx600-1000$, and the DM annihilation peaks at $z\approx600$ \cite{Planck:2015fie}. The constraint also applies to the AQN annihilation as discussed in section \ref{subsec:Anisotropies from AQNs}.

\subsection{AQN-induced signatures}
\label{sec:AQN-induced signatures}
The explicit CMB signatures on both the spectral distortions and anisotropies can be computed using \texttt{CLASS} by inputting the heat injection $\dot{Q}$ and the cross section $\sigma$ from eqs. \eqref{eq:dot Q} and \eqref{eq:sigma pi R 2}. Without sophisticated numerical simulation, several key features of the AQN-induced signatures can still be extracted from the approximate heat injection $\dot{Q}_{\rm approx}$ from eq. \eqref{eq:dot Q_apprx}.

In this section, we will obtain several expected features of the spectral distortions and anisotropies from AQN energy injection based on simple analysis. The expected features will be compared with the numerical calculations presented in section \ref{sec:Results}.

\subsubsection{Spectral distortion from AQNs}
Spectral distortions in the CMB are commonly characterized by the parameters $\mu$ and $y$, which quantify fractional deviations in the photon energy and number density from a perfect blackbody distribution. Their approximate forms, simplified from Eq. \eqref{eq:Delta I_tot nu z}, are given by (see e.g. \cite{Chluba:2013vsa,Chluba:2016bvg,Chluba:2018cww}):
\begin{subequations}
\label{eqs:y etc.}
\begin{equation}
y
=\left.\frac{1}{4}\frac{\Delta\rho_\gamma}{\rho_\gamma}\right|_y
=\frac{1}{4}c^{-2}
\int_0^\infty{\cal J}_y(z')\frac{\rmd(Q/\rho_\gamma)}{\rmd z'}\rmd z'\,;
\end{equation}
\begin{equation}
\mu
=1.401\left.\frac{\Delta\rho_\gamma}{\rho_\gamma}\right|_\mu
=1.401c^{-2}
\int_0^\infty{\cal J}_\mu(z')\frac{\rmd(Q/\rho_\gamma)}{\rmd z'}\rmd z'\,,
\end{equation}
\end{subequations}
with the distortion visibility
\begin{subequations}
\label{eqs:cal J_y z etc.}
\begin{equation}
{\cal J}_y(z)
\approx\left\{\begin{aligned}
&\left[1+\left(\frac{1+z}{6\times10^4}\right)^{2.58}\right]^{-1}\,,&
z\gtrsim 10^3  \\
&0\,,&{\rm\,otherwise}
\end{aligned}\right.;
\end{equation}
\begin{equation}
{\cal J}_\mu(z)
\approx {\cal J}_{\rm bb}(z)\left\{
1-\exp\left[-\left(\frac{1+z}{5.8\times10^4}\right)^{1.88}\right]
\right\}\,,\quad
{\cal J}_{\rm bb}(z)\approx\exp\left[-\left(\frac{z}{1.98\times10^6}\right)^{2.5}\right]\,.
\end{equation}
\end{subequations}

The $y$-distortion typically arises from heating events at redshifts $z\lesssim10^4$, while the $\mu$-distortions originate from earlier epochs, $10^5\lesssim z\lesssim 2\times 10^6$, when thermalization becomes inefficient. The heating mechanism can be verified from the eqs. \eqref{eqs:y etc.} and \eqref{eqs:cal J_y z etc.}.The distortion visibilities ${\cal J}_y(z)$ and ${\cal J}_\mu(z)$ are approximately flat and of order one near the redshifts $z\sim5\times10^4$ and $10^6$, respectively, which implies:

\begin{equation}
\label{eq:y_mu_estimate}
y 
\sim \frac{1}{4}c^{-2}
\left. \frac{\Delta Q}{\rho_\gamma} \right|_{z=5\times10^4}\,, \qquad
\mu 
\sim 1.401c^{-2}\left. \frac{\Delta Q}{\rho_\gamma} \right|_{z=10^6}\,,
\end{equation}
where $\Delta Q$ is the total heat injected into the photon bath in the redshift range $[z-\frac{1}{2}\Delta z,z+\frac{1}{2}\Delta z]$:
\begin{equation}
\Delta Q 
\equiv \int_{z-\frac{1}{2}\Delta z}^{z+\frac{1}{2}\Delta z}
\frac{\rmd Q}{\rmd z'}\rmd z' \approx \Delta z \frac{\mathrm{d}Q}{\mathrm{d}z}\,.
\end{equation}
Based on the full width at half maximum of the distortion visibilities \eqref{eqs:cal J_y z etc.}, we find $\Delta z\approx5\times10^4$ for $y$-distortion and $\Delta z\approx10^6$ for $\mu$-distortion. These values coincide with the choice of $z$'s in eqs. \eqref{eq:y_mu_estimate}. Thus, the width of the redshift will be estimated as $\Delta z\approx z$, the total injected energy $\Delta Q$ can be related to $\dot{Q}$ via:
\begin{equation}
\Delta Q 
\approx \frac{\Delta z}{1+z} \frac{\dot{Q}}{H(z)}
\approx\frac{\dot{Q}}{H(z)}\,,
\end{equation}
where $H(z)$ is the Hubble expansion rate. In the radiation-dominated era, we have:
\begin{equation}
H(z) \approx H_0\Omega_{\gamma,0}^{1/2}(1+z)^2\,,\qquad
\rho_\gamma = \rho_{\gamma,0}(1+z)^4\,,
\end{equation}
where $H_0$ is the present day Hubble expansion rate, $\rho_{\gamma,0}$ is the present day radiation density, and $\Omega_{\gamma,0}$ is the present day radiation density parameter.

The spectral distortion \eqref{eq:y_mu_estimate} suggests that $y,\mu\propto\Delta Q/\rho_\gamma$, which can be estimated as follows:
\begin{equation}
\label{eq:y mu}
\begin{aligned}
y,\mu
\sim 
\frac{\Delta Q}{\rho_\gamma} \approx \frac{1}{H_0 \Omega_{\gamma,0}^{1/2}\rho_{\gamma,0}} \frac{\dot{Q}}{(1+z)^6}
=\frac{8\pi G_{\rm N}}{3H_0^3\Omega_{\gamma,0}^{3/2}}
\frac{\dot{Q}}{(1+z)^6}\,, 
\end{aligned}
\end{equation}
where $G_{\rm N}$ is Newton's gravitational constant. Assuming the ionized hydrogen is the dominant interacting species, i.e. $\sigma\approx \pi R_{\rm eff}$ from eqs. \eqref{eq:R_eff R 2} and \eqref{eq:sigma pi R 2}, we obtain corresponding $y$ and $\mu$ distortion parameters :
\begin{equation}
\label{eq:Delta Q rho_gamma ::AQN}
\begin{aligned}
c^{-2}\frac{\Delta Q}{\rho_\gamma}
\approx\frac{8\pi G_{\rm N}}{3c^2H_0^3\Omega_{\gamma,0}^{3/2}}
\frac{\dot{Q}_{\rm approx}}{(1+z)^6}
=4.33\times10^{-11}(1+z)^{1/2}
\left(\frac{g}{0.1}\right)
\left(\frac{\rm\Delta v}{10^{-4}c}\right)\,.\quad
{\rm\,(AQN)}
\end{aligned}
\end{equation}
Note that the distortion has no mass dependence on $m_{\rm AQN}$, and we neglect the redshift dependence of $\rm\Delta v$ for simplicity. Applying the approximation \eqref{eq:y_mu_estimate}, we estimate:
\begin{equation}
\label{eqs:y sim etc.}
y\sim2\times10^{-9}\,,\qquad
\mu\sim6\times10^{-8}\,.\qquad
{\rm\,(AQN)}
\end{equation}
These values would meet the PIXIE detection limit of approximately $10^{-8}$, as will be discussed in section \ref{sec:Discussion and Conclusion}.

Notably, the AQN signature has a higher $\mu$-distortion compared to its $y$-distortion. This is nontrivial in conventional DM models. To see the difference, we formulate the general form of energy injection rate for DM candidates according to eqs. \eqref{eqs:dot Q_ann etc}:
\begin{equation}
\dot{Q}\propto (1+z)^n\,,
\end{equation}
where $n\approx3$ for DM decay, and $n\approx6$ for DM annihilation. From equation \eqref{eq:y mu}, the corresponding $y$-to-$\mu$ ratio is:
\begin{subequations}
\begin{equation}
\left(\frac{y}{\mu}\right)_{\rm dec}
\sim0.178\left.\left(\frac{10^6}{5\times10^4}\right)^{6-n}\right|_{n\approx3}
\gg1\,,\qquad
{\rm\,(DM~decay)}
\end{equation}
\begin{equation}
\left(\frac{y}{\mu}\right)_{\rm ann}
\sim0.178\left.\left(\frac{10^6}{5\times10^4}\right)^{6-n}\right|_{n\approx6}
\sim 0.1-1\,,\qquad
{\rm\,(DM~annihilation)}
\end{equation}
\end{subequations}
where the prefactor 0.178 comes from the ratio of $\frac{1}{4}$ and 1.401 in formulae \eqref{eq:y_mu_estimate}.
The simple estimate indicates that the $y$-distortion is either essentially larger than the $\mu$-distortion (DM decay) or comparable to the $\mu$-distortion (DM annihilation). In conventional DM models, including macroscopic scenarios such as those in \cite{Kumar:2018rlf}, $\mu$-distortions rarely exceed $y$-distortions by an order of magnitude. However, this does not hold for AQNs, for which the opposite is true. We will verify this conclusion in the next section by performing explicit numerical simulations. One can trace that this feature is an explicit manifestation of the unique property of the AQN model when the energy injection rate is proportional to the baryonic and DM densities, i.e. $\propto (n_{\rm DM}\cdot n_{\rm b})$ as discussed after (\ref{eq:Q_0}), in contrast with canonical approaches when it is proportional to $n_{\rm DM}$ (decaying)  or $n_{\rm DM}^2$ (annihilating) DM densities. The $\mu$ and $y$ amplitude difference between DM annihilation, DM decay, and AQN can be understood as follows:
We found that $(y/\mu)\propto(z_\mu/z_y)^{6-n}$, where $z_\mu\approx10^6$ and $z_y\approx5\times10^4$ are the corresponding redshift range for $\mu$- and $y$-distortion. Specifically, $n=3,6$ and 6.5 for DM decay, DM annihilation, and AQNs, which explains the amplitude differences between the three DM models.

\subsubsection{Anisotropies from AQNs}
\label{subsec:Anisotropies from AQNs}

The energy injection of AQNs is similar to that of the DM annihilation. Comparing eq. \eqref{eq:dot Q} with eqs. \eqref{eq:dot Q_ann} and \eqref{eq:p_ann}, we find that the equivalent $p_{\rm ann}$ for the AQN model is:
\begin{equation}
\begin{aligned}
p_{\rm ann,AQN}
&\approx\left.\frac{2}{5}
\frac{E_{\rm ann}}{m_pc^2}\frac{\Omega_{\rm b,0}}{\Omega_{\rm DM,0}}
\frac{g\sigma {\rm\Delta v}}{m_{\rm AQN}c^2}\right|_{z=600}
\approx\left.\frac{4}{5}
\frac{\Omega_{\rm b,0}}{\Omega_{\rm DM,0}}
\frac{g\sigma {\rm\Delta v}}{m_{\rm AQN}c^2}\right|_{z=600}
\\
\end{aligned}
\end{equation}
where we approximate $E_{\rm ann}\approx2m_pc^2$ for simplicity; $\Omega_{\rm b,0}$ and $\Omega_{\rm DM,0}$ are the baryon and DM density parameters.
At redshift $z\simeq 600$, the Universe is almost neutral, and the ionization fraction is small, $X_e\lesssim10^{-3}$ \cite{Padmanabhan:2005es,Finkbeiner:2011dx}. From eq. \eqref{eq:sigma pi R 2}, the AQN weighted cross section is of the same order as the geometrical cross-section$\sigma(z=600)=\pi R^2$. This implies:
\begin{equation}
\label{eq:p_ann AQN}
\begin{aligned}
&p_{\rm ann,AQN}
\lesssim4.18\times10^{-30}{\rm\,cm^3s^{-1}GeV^{-1}}
\left(\frac{100{\rm\,g}}{m_{\rm AQN}}\right)^{\frac{1}{3}}
\left(\frac{g}{0.1}\right)
\left(\frac{\rm\Delta v}{10^{-4}c}\right)
\left(\frac{\sigma(z=600)}{\pi R^2}\right)\,.
\end{aligned}
\end{equation}
The estimate suggests that the energy injection of the AQNs is similar to the conventional DM annihilation with $p_{\rm ann}\lesssim10^{-29}{\rm\,cm^3s^{-1}GeV^{-1}}$. The AQN model does not violate the constraint \eqref{eq:p_ann} on CMB anisotropies observed from Planck. The conclusion is insensitive to the mass of AQNs because $p_{\rm ann,AQN}\sim m_{\rm AQN}^{-1/3}$. As we will see in the following section, the simple analytical estimation \eqref{eq:p_ann AQN} agrees well with the explicit numerical simulation.
\footnote{Note that the last term $\frac{\sigma(z=600)}{\pi R^2}$ in eq. \eqref{eq:p_ann AQN} is insensitive to $m_{\rm AQN}$ because both $\sigma(z=600)$ and $\pi R^2$ scale like $R^2\sim m_{\rm AQN}^{2/3}$. Their ratio cancels the dependence on $m_{\rm AQN}$.}

\begin{figure}[h!]
\centering
\includegraphics[width=0.9\textwidth]{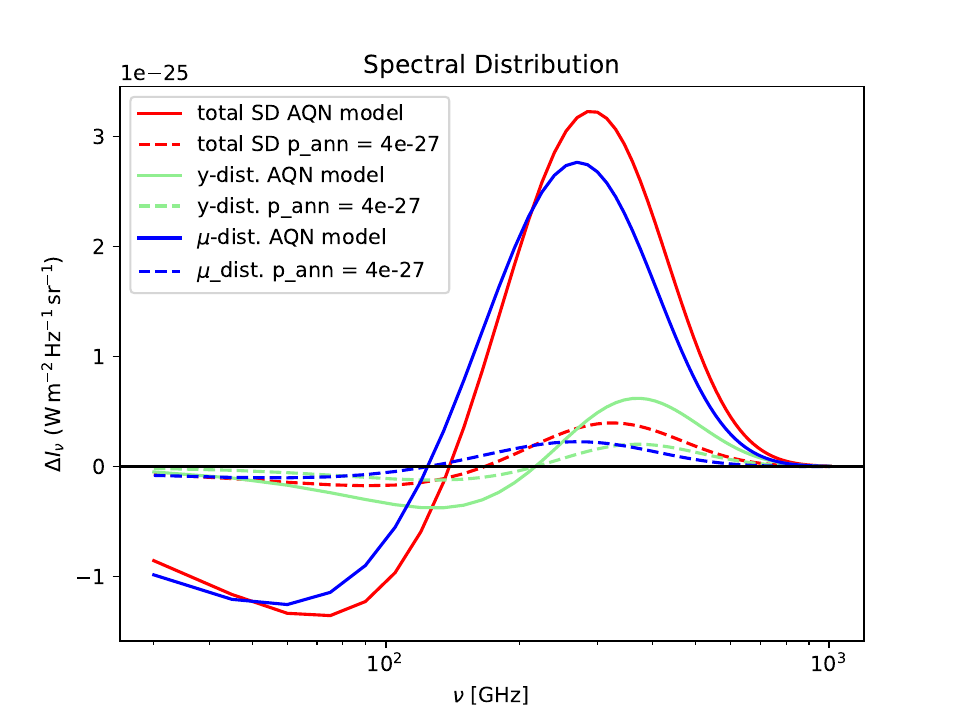}
\caption{Spectral distortion ($\Delta I_\nu$) contributions from Axion Quark Nugget (AQN) models compared to a dark matter annihilation model with $p_{\rm ann} = 4 \times 10^{-27}~\mathrm{cm^3\,s^{-1}\,GeV^{-1}}$. While the annihilation scenario exceeds Planck's upper bound and is observationally ruled out, its resulting distortions remain significantly smaller than those from AQN models. This illustrates that even observationally excluded levels of dark matter annihilation lead to weaker spectral distortion signatures than physically viable AQN scenarios, underlining the strong thermal visibility of AQN-induced distortions.}
\label{fig:SD_AQN_Ann}
\end{figure}

\section{Results}
\label{sec:Results}

We present the spectral-distortion signatures predicted by the AQN model and compare them with current observational constraints and the projected sensitivities of future CMB spectral-distortion experiments. In particular, we compare the predicted $\mu$ and $y$-type distortions with the upper limits established by COBE/FIRAS \cite{Fixsen:1996nj} and with the expected sensitivities of forthcoming missions such as Voyage 2050 \cite{2021ExA....51.1515C}, PIXIE \cite{2011JCAP...07..025K,2020JCAP...05..041K,Kogut:2024vbi}.

\subsection{Spectral Distortion Predictions from AQN Model}

Figure \ref{fig:SD_AQN_Ann} shows the spectral distortions predicted by the AQN model for the average nugget mass of $10\;\rm g$. The curves display the total distortion as well as the individual $\mu$ and $y$ components. For comparison, we also include the distortion produced by dark-matter annihilation with a $p_{\rm ann}$ parameter exceeding the Planck upper limit by a factor of $\sim10$. As the figure illustrates, the AQN-induced distortions are substantially larger than those expected from conventional annihilation scenarios, even in this upper-limit case, highlighting the strong observational prospects for distinguishing an AQN signal. We note that the Planck constraint on $p_{\rm ann}$ is driven by the excessive optical depth generated by DM–DM annihilation at $z<1000$ and not by the FIRAS measurements. As discussed in section \ref{sec:ani}, the AQN model does not suffer from this issue.

\begin{figure}[htbp]
\centering
\includegraphics[width=0.48\textwidth]{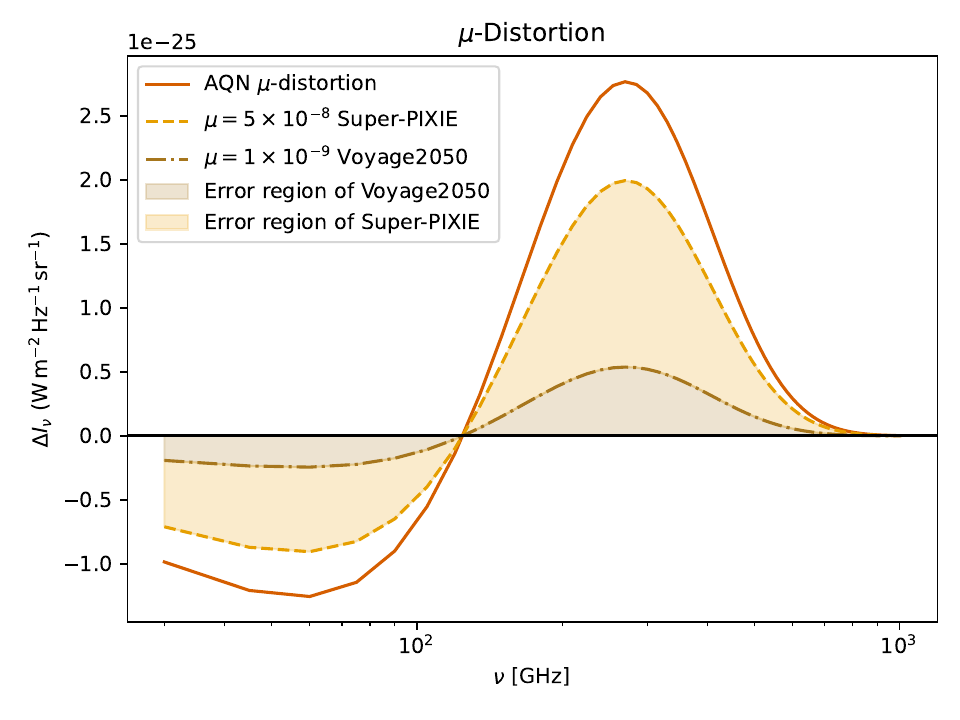}
\includegraphics[width=0.48\textwidth]{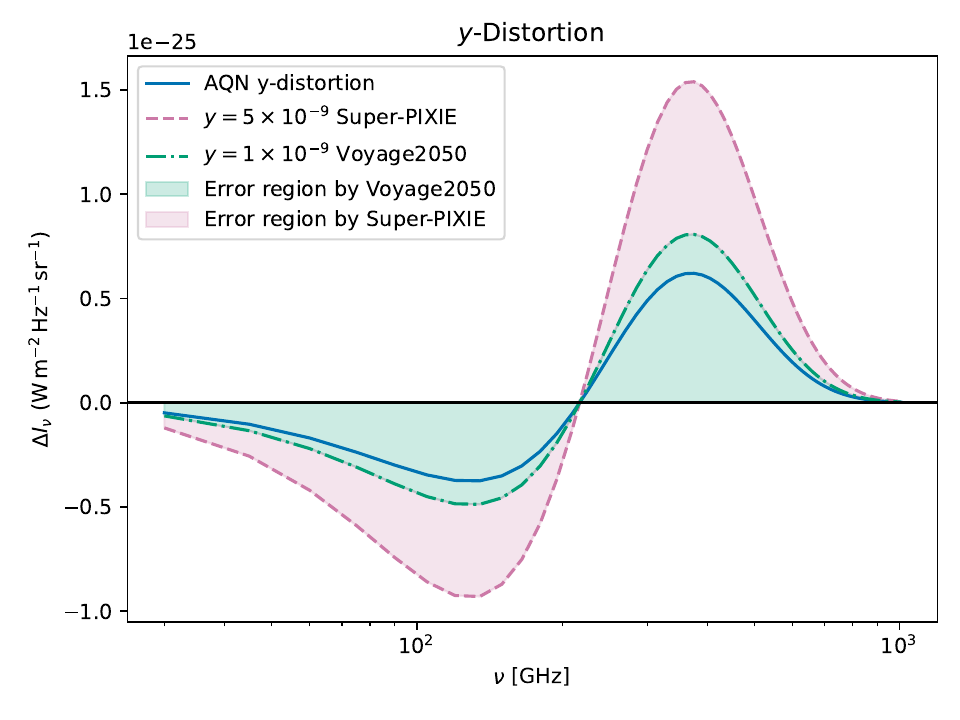}
\caption{Spectral distortions in the CMB predicted by the AQN model. Left: Projected sensitivities to $\mu$-type distortions. The solid red line shows the $\mu$-distortion spectrum from the AQN model, while the dashed and dash-dotted lines represent the expected detection thresholds of PIXIE ($\mu=5\times10^{-8}$) and Voyage2050 ($\mu=1\times10^{-9}$), respectively.
Right: Predicted $y$-type distortion signal from the AQN model (solid blue), shown in comparison to the projected sensitivities of PIXIE ($y=5\times10^{-9}$, dashed green) and Voyage2050 ($y=1\times10^{-9}$, dash-dotted magenta).}
\label{fig:SD_limits}
\end{figure}

Can the AQN signal be detected by proposed mission concepts such as Super-PIXIE or Voyage 2050? These designs target sensitivities of approximately $\mu \sim 5\times 10^{-8}$ and $y \sim 5 \times 10^{-9}$ for super-PIXIE and $\mu \sim y \sim 10^{-9}$ for Voyage 2050.
Figure \ref{fig:SD_limits} compares the predicted AQN signal with these projected performance levels. Each mission’s sensitivity is represented by a solid coloured band across frequency space, while the vertical axis shows the spectral distortion relative to a blackbody in physical units ($\rm W m^{-2}Hz^{-1}sr^{-1}$).
The strong $\mu$ type distortion exceeds the $1\sigma$ sensitivity across all frequencies, implying that it would be detectable by both mission concepts, even with only a limited number of frequency channels. In contrast, the $y$-type distortion lies below the $1\sigma$ threshold at all frequencies, but remains detectable provided a sufficient number of channels are available. Notably, the much larger amplitude of the $\mu$ distortion relative to the $y$ distortion is a distinctive feature of the AQN model, enabling its detection through the $\mu$ channel even when the $y$ channel alone would be insufficient.

A potential source of confusion is the Sunyaev–Zeldovich (SZ) spectral distortion. 
However, the AQN-induced distortion is observationally distinguishable from the SZ effect. 
The SZ distortion is a local, low-redshift phenomenon arising from inverse Compton 
scattering of CMB photons by hot electrons in collapsed structures such as galaxy clusters. 
It therefore produces spatially localized distortions that can be identified and masked in 
high-resolution surveys. In contrast, the AQN-induced distortion originates from energy 
injection at high redshifts in the early Universe and contributes to the global (monopole) 
component of the CMB spectral distortion. The two signals thus differ both in their redshift 
origin and spatial characteristics, making them distinguishable in spectral-distortion 
measurements.

We emphasize that Fig.~2 is intended only as an order-of-magnitude comparison between the predicted AQN-induced spectral distortion and the nominal sensitivities of future experiments. A quantitative assessment of detectability would require a dedicated forecast including instrumental noise, frequency coverage, foreground modeling, and component separation. In particular, the ability to isolate a specific spectral signature depends on component separation techniques, which may, in principle, allow sensitivity to sub-$\sigma$ signals. A full analysis of this kind is beyond the scope of the present work.

\subsection{CMB Anisotropy and Cosmological Parameters}
\label{sec:ani}

\begin{figure}
    \centering
    \includegraphics[width=0.9\linewidth]{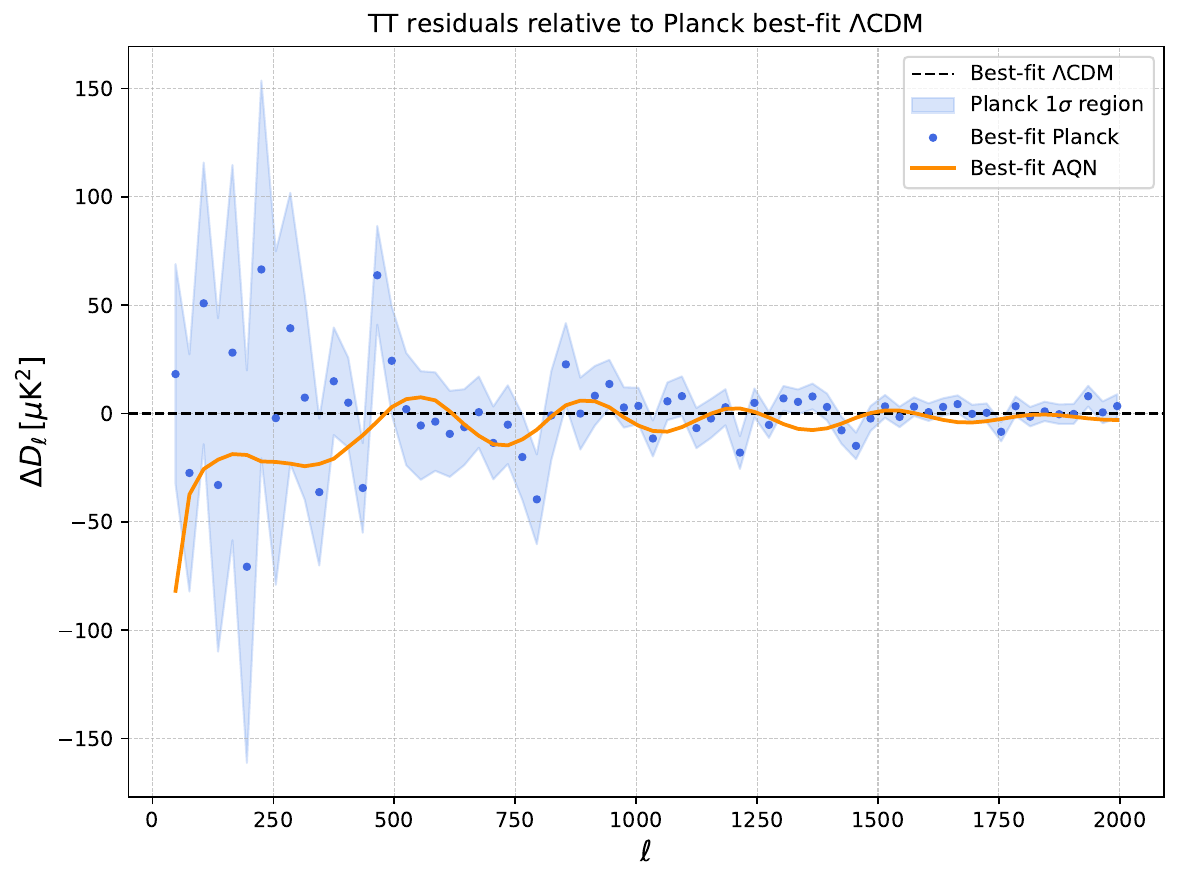}
    \caption{Residual TT angular power spectrum relative to the Planck best-fit $\Lambda$CDM model. The horizontal dashed line at $\Delta D_\ell = 0$ corresponds to the Planck best-fit $\Lambda$CDM spectrum. The blue points show the residuals of the Planck TT binned data relative to the Planck best-fit $\Lambda$CDM model, while the shaded blue region represents the associated Planck $1\sigma$ uncertainties. The solid orange curve shows the residuals of the best-fit AQN model with respect to the same Planck best-fit $\Lambda$CDM spectrum for $m_{\rm AQN}=10 {\rm g}$. The figure demonstrates that the AQN-induced modifications remain within the observational uncertainties of the Planck TT measurements over the multipole range shown.}
    \label{fig:TT_AQN}
\end{figure}

\begin{figure}
    \centering
    \includegraphics[width=1\linewidth]{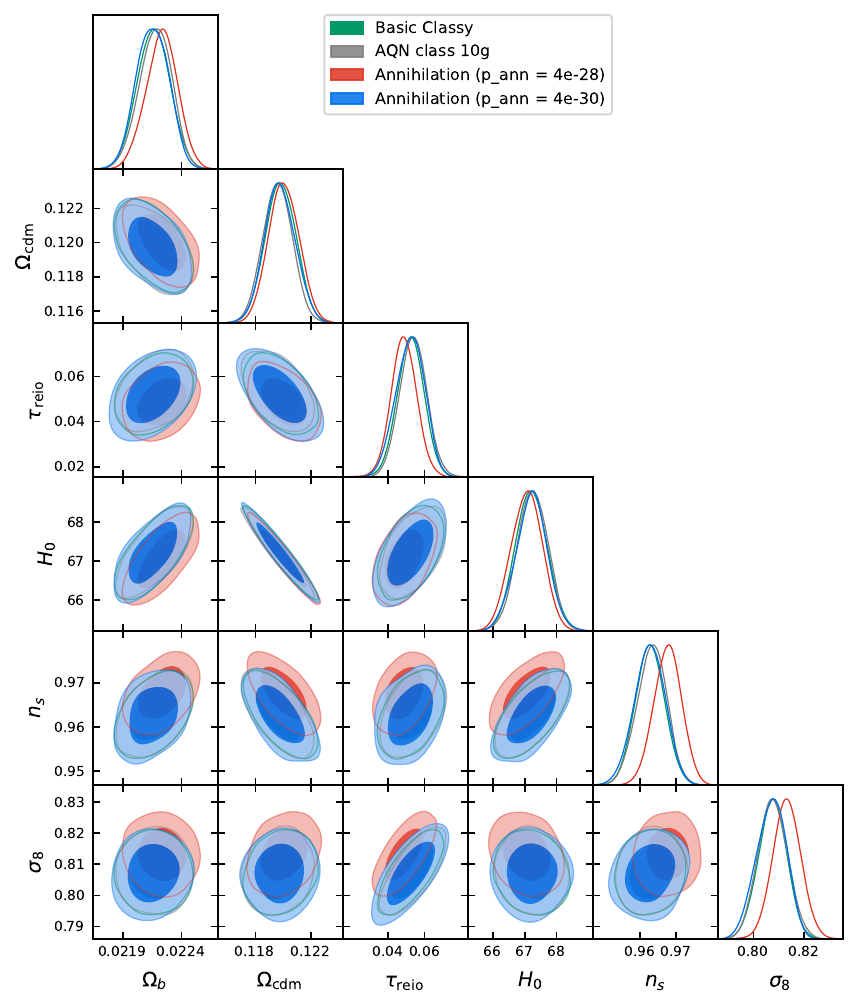}
    \caption{Corner plots showing the posterior distributions for six cosmological parameters under four different models: the standard \textit{Basic Classy} run, the AQN model with an average mass of 10g, and two dark matter annihilation scenarios with $p_{\rm ann} = 4 \times 10^{-28}~\mathrm{cm^3\,s^{-1}\,GeV^{-1}}$ and $p_{\rm ann} = 4 \times 10^{-30}~\mathrm{cm^3\,s^{-1}\,GeV^{-1}}$. Among these, only the red curve corresponding to the $p_{\rm ann} = 4 \times 10^{-28}$ $\mathrm{cm^3\,s^{-1}\,GeV^{-1}}$ case exceeds the observational upper bound from Planck, while the others remain within the allowed region.}
    \label{fig:Triangle_plot} 
\end{figure}

While the AQN model generates distinctive spectral-distortion signatures, its impact on the primary CMB anisotropies is minimal. Figure \ref{fig:SD_AQN_Ann} showed that the spectral-distortion signal produced by AQNs is far stronger than that expected from dark-matter annihilation with a rate already excluded by Planck. This raises the question of whether AQNs could also produce a significant anisotropy signal. To address this, we used our modified version of \texttt{CLASS} to forecast cosmological parameter constraints for an AQN model similar to the one used in Figure \ref{fig:SD_limits}, and compared them with the constraints obtained from an annihilating dark-matter model that is consistent with the Planck detection limit.

The results are shown in Figure \ref{fig:Triangle_plot}, which presents corner plots comparing the posterior distributions of six key cosmological parameters under four different scenarios: the standard \textit{Basic Classy} $\Lambda$CDM run (green), the AQN model with an average nugget mass of 10g (grey), and two dark matter annihilation models with energy injection rates of $p_{\rm ann} = 4 \times 10^{-28}~\mathrm{cm^3\,s^{-1}\,GeV^{-1}}$ (red) and $p_{\rm ann} = 4 \times 10^{-30}~\mathrm{cm^3\,s^{-1}\,GeV^{-1}}$ (blue), respectively. For this analysis, we employed the Cobaya framework \cite{Torrado:2020dgo} and used a combination of high- and low-$\ell$ likelihoods from Planck 2018 and Planck PR4. Specifically, we included the low-$\ell$ temperature and polarization likelihoods \textit{planck\_2018\_lowl.TT} and \textit{planck\_2018\_lowl.EE} \cite{Planck:2019nip}, the high-$\ell$ CamSpec temperature and polarization likelihood \textit{planck\_NPIPE\_highl\_CamSpec.TTTEEE} \cite{Planck:2019nip}, and the Planck PR4 lensing likelihood \textit{planckpr4lensing} \cite{Carron:2022eum, Carron:2022eyg}, installed via the public GitHub repository.

As shown in Fig.~\ref{fig:Triangle_plot}, the AQN-induced energy injection leads to
only negligible shifts in the posterior distributions of
$\Omega_{\rm b} h^2$, $\Omega_{\rm cdm} h^2$, $H_0$, $n_s$, $\tau_{\rm reio}$, and $\sigma_8$.
The posterior contours for the AQN model (grey) almost completely overlap with those of
the base $\Lambda$CDM case (green), indicating that the AQN model remains fully consistent with the \textit{Planck} data.
For our fiducial AQN scenario with an average nugget mass
$\langle m_{\rm AQN}\rangle = 10\,\mathrm{g}$, we obtain a reduced chi-squared
$\chi^2_{\rm red}=1.09$, demonstrating an acceptable goodness-of-fit and no degradation
relative to the base $\Lambda$CDM model.


To better illustrate the impact of the AQN model on the CMB TT power spectrum, in Fig. \ref{fig:TT_AQN} we show the residuals relative to the Planck best-fit $\Lambda$CDM model. The horizontal line at $\Delta D_\ell = 0$ corresponds to the Planck best-fit $\Lambda$CDM spectrum. The blue points show the residuals of the Planck TT binned data relative to the same best-fit $\Lambda$CDM model, while the shaded region represents the corresponding Planck $1\sigma$ uncertainty range. The AQN curve shows the difference between the best-fit AQN model and the Planck best-fit $\Lambda$CDM spectrum. This representation allows a direct visualization of the magnitude and multipole dependence of the AQN-induced modification compared to the observational uncertainties of the Planck TT measurements.

By contrast, the illustrative annihilation scenario with a fixed value
$p_{\rm ann}=4\times10^{-28}\,\mathrm{cm^3\,s^{-1}\,GeV^{-1}}$ (red) induces visible shifts
in parameters such as $\tau_{\rm reio}$ and $\sigma_8$ along known degeneracy directions.
These shifts should not be interpreted as internal inconsistencies of the extended
cosmological model; rather, they reflect the fact that this annihilation strength exceeds
the direct \textit{Planck} constraints on dark-matter annihilation.
Indeed, \textit{Planck} CMB anisotropy measurements place 95\% confidence-level upper
limits on the annihilation parameter $p_{\rm ann}$ at the level of a few
$\times10^{-28}\,\mathrm{cm^3\,s^{-1}\,GeV^{-1}}$, thereby disfavouring such aggressive
annihilation scenarios (see Sec.~6.6 of \cite{Planck:2015fie}).

These results affirm that, unlike certain exotic dark matter models, AQN-induced heating remains safely within current observational constraints, which is consistent with the analytical estimation \eqref{eq:p_ann AQN}. Consequently, the standard cosmological framework retains its robustness, even under scenarios involving additional physics such as AQNs. This highlights the viability of AQNs as an exotic dark matter candidate that introduces detectable spectral features while preserving the integrity of the CMB angular power spectrum. We restrict the comparison to multipoles $\ell \lesssim 2000$, where the CMB signal is dominated by primary anisotropies. At higher multipoles, foreground contributions and secondary anisotropies become dominant (see, e.g., \cite{SPT2021} figure 2), and a consistent analysis would require detailed foreground modeling, which is not included in the present analysis

\begin{figure}
    \centering
    \includegraphics[width=1\linewidth]{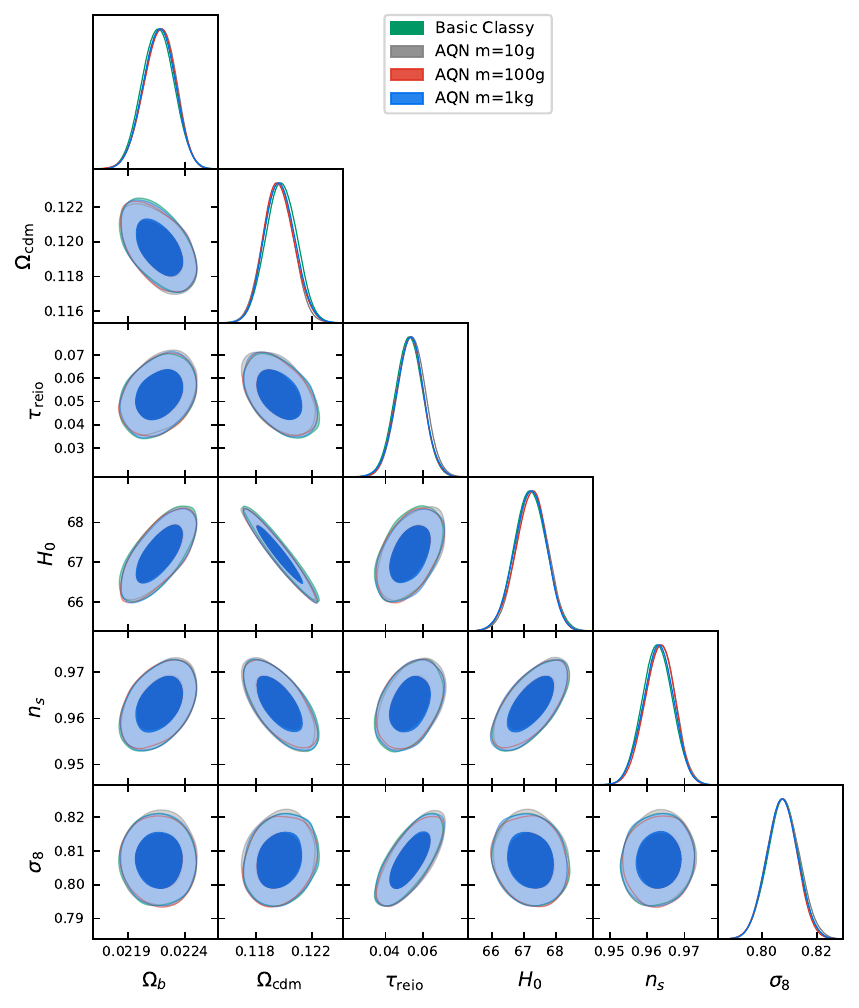}
    \caption{Corner plots showing the posterior distributions for six cosmological parameters under three scenarios: the standard \textit{Basic Classy} model, and three Axion Quark Nugget (AQN) models with nugget average masses of 10 g and 100 g, and 1 kg. All cases remain consistent with the Planck upper limit on $p_{\rm ann}$, supporting the viability of AQN dark matter within current observational bounds.}
    \label{fig:TPlot_mass}
\end{figure}

Moreover, as illustrated in figure~\ref{fig:TPlot_mass}, we examined the effect of varying the average mass of Axion Quark Nuggets (AQNs) over a wide range—from $10\,\mathrm{g}$ to $1000\,\mathrm{g}$—on cosmological parameter constraints. The corner plots show that such variations do not induce any significant shifts in key cosmological parameters such as $\Omega_\mathrm{cdm}$, $H_0$, $n_s$, $\sigma_8$, or $\tau_\mathrm{reio}$. All contours for different AQN masses remain virtually indistinguishable and tightly overlap with the baseline $\Lambda$CDM contours. This confirms that the AQN heat injection remains subdominant and does not alter the consistency of the $\Lambda$CDM model with current CMB anisotropy measurements. This is an expected feature from estimation \eqref{eq:p_ann AQN} because $p_{\rm ann,AQN}\sim m_{\rm AQN}^{-1/3}$.

It is important to note that in the AQN model, the only free parameter is the average mass of the nuggets. The negligible impact observed across this mass range demonstrates that changes in AQN mass do not meaningfully affect the thermal history or the derived cosmological parameters. However, in a separate study \cite{Majidi:2024mty}, we have shown that it is still possible to place meaningful upper limits on the AQN mass using constraints from CMB spectral distortions. This dual conclusion highlights the robustness of the AQN model: while its cosmological imprints in terms of parameter shifts are small, it remains a testable and distinguishable scenario via high-precision spectral observations.

\section{Discussion and Conclusion}
\label{sec:Discussion and Conclusion}

In this work, we investigated the spectral distortion signatures predicted by the AQN dark matter model and compared them against current observational bounds and the expected sensitivities of future CMB missions. Our calculations show that the AQN model produces $\mu$-type distortions well within the projected detection capabilities of next-generation experiments such as PIXIE and Voyage 2050. While the associated $y$-type distortions remain below the $1\sigma$ sensitivity thresholds, their spectral shape is robust and consistent, indicating that with improved frequency coverage or reduced instrumental noise, they could still be probed in future datasets. This highlights the potential of spectral distortion measurements as a powerful probe for AQN-induced energy injection, even in regimes where full detection remains challenging.

Furthermore, we explored the impact of AQNs on the primary CMB anisotropies and cosmological parameters by conducting a series of Markov chain Monte Carlo (MCMC) analyses We found that the inclusion of AQN-induced heating does not lead to any significant deviations in key cosmological parameters such as $\Omega_{\rm b} h^2$, $\Omega_{\rm cdm} h^2$, $H_0$, $n_s$, $\tau_{\rm reio}$, and $\sigma_8$. The posterior distributions for the AQN scenarios remain tightly consistent with the baseline $\Lambda$CDM model, demonstrating that the AQN contributions are cosmologically safe. We also tested the sensitivity of these results to the average mass of the nuggets—currently the only free parameter in the AQN model—and found that varying the mass across a broad range from 10 g to 1000 g has no observable effect on the cosmological constraints. 

One should emphasize that our choice for $m_{\rm AQN}\in (10, 100)$ g is not an arbitrary selection for an allowed window. Instead, the lower bound appears from direct (non)observation by the IceCube collaboration; see the appendix in \cite{Lawson:2019cvy}. The upper bound emerges from the study~\cite{Majidi:2024mty}, where we have shown that one can derive an upper limit on the AQN mass using the spectral distortion signatures alone.  In addition,  the AQN-induced signal  in UV frequency bands  \cite{Sekatchev:2025ixu}     
may explain the excess of the UV emission which has been observed \cite{Henry_2014,Murthy_2025}. The AQN mass in the computations \cite{Sekatchev:2025ixu} assumes the values within the same window  $m_{\rm AQN}\in (10, 100)$ g. 

The same allowed window also appears in our previous studies (with dramatically different scales and environments), such as the primordial Lithium Puzzle \cite{Flambaum:2018ohm} or solar corona heating puzzle \cite{Raza:2018gpb}, when the AQN-induced processes could be responsible for the resolution of the corresponding long-standing problems. We consider this feature as a highly nontrivial element of the AQN construction supporting the self-consistency of the entire approach when applications of the AQN model to very different problems with dramatically different scales nevertheless lead to the same allowed window $m_{\rm AQN}\in (10, 100)$ g.
These results emphasize that the AQN model provides a consistent and observationally viable extension to the standard cosmological framework, with distinct and testable features accessible through future high-precision CMB experiments.

\subsection{Future Work}

\subsubsection{Low-energy Photons}

In this work, we explored the cosmological implications of the AQN dark matter model by focusing on its contribution to the spectral distortions of the CMB and the resulting constraints from current and future CMB observations. Our analysis concentrated primarily on the high-energy (above $\sim 1{\rm\,keV}$) photon emission from AQNs and demonstrated that the predicted $\mu$- and $y$-type distortions lie well below the COBE/FIRAS limits, while remaining potentially detectable by proposed experiments such as PIXIE, FOSSIL, and Voyage 2050.

However, one natural and important extension of this work lies in studying the low-energy photon emission from AQNs. The thermal emmission of AQNs is very close to a bremsstrahlung spectrum covering a wide range of frequencies (from GHz to UV). For wavelengths at the Ly$\alpha$ and below, there are a number of atomic transitions which can play a very important role in the history of the ionization parameter $X_e$ before and after recombination, the 21 cm spin temperature, in particular during the cosmic dawn era. 

Although these processes are less likely to modify the $\mu$ and $y$-type distortions, they can leave subtle, yet detectable, imprints in the Rayleigh-Jeans tail of the sky spectrum. Additionally, the anisotropies of these low-frequency SDs could provide new insights into the spatial distribution and clustering properties of AQNs.

While our current treatment of the AQN energy injection at high redshift focused on the high-energy end of the AQN emission spectrum using analytic approximations, future studies should incorporate the full low-energy spectrum by employing advanced numerical tools such as \texttt{CosmoTherm} \cite{Chluba:2011hw}. This Boltzmann code allows precise modelling of energy injection and thermalization processes across the entire photon energy spectrum taking into account atomic transitions. By feeding the detailed low-energy photon production spectrum into \texttt{CosmoTherm}, it will be possible to generate more accurate predictions of the resulting distortions and anisotropies in this regime.

Moreover, studying the low-energy contribution is particularly timely given the design of upcoming radio-band SD experiments, such as the proposed low-frequency extension of the PIXIE mission \cite{2011JCAP...07..025K,2020JCAP...05..041K,Kogut:2024vbi} and ground-based low-frequency receivers \cite{Mukherjee:2018fxd, Abitbol:2017vwa}. These instruments may offer complementary sensitivity to the soft photon tail, potentially constraining AQN-induced radiation in a region of the spectrum not yet explored observationally. Such work could also help discriminate between different dark matter candidates that produce low-frequency signatures.

In conclusion, while the high-energy spectral distortions from AQNs serve as a compelling and detectable signal for future CMB experiments, a comprehensive understanding of the full emission spectrum—including the low-energy tail—will provide a more complete observational test of the AQN model. This direction remains an exciting and feasible avenue for future research.

\subsubsection{Synergy with BISOU and Balloon-Borne SD Observations}

In light of ongoing and upcoming efforts to detect CMB spectral distortions, the BISOU project~\cite{bisou} offers a particularly promising avenue. As a French balloon-borne mission designed to measure absolute CMB spectra at high sensitivity across the 70–400~GHz range, BISOU aims to detect $\mu$-type distortions at the $10^{-7}$ level—potentially bridging the observational gap between COBE/FIRAS and future space-based missions like PIXIE or Voyage 2050.

Although our AQN model primarily predicts spectral distortions that lie below the current COBE/FIRAS upper bounds (with typical values $y \sim 10^{-6}$–$10^{-7}$ and $\mu \lesssim 10^{-8}$), BISOU’s design sensitivity could already begin probing the parameter space relevant to AQN-induced energy injection. The balloon-based frequency coverage of BISOU overlaps significantly with the peak emission regions of AQN spectral templates, particularly in the GHz to sub-THz range.

Furthermore, BISOU's strategy of high-altitude, short-duration flights enables complementary data points across atmospheric windows that are difficult to access from ground-based observatories. This is particularly important in the context of AQN models, where distortions may have subtle frequency-dependent features that can be better disentangled with multi-frequency sampling. Even if the baseline AQN parameters are not within immediate reach, stacking multiple BISOU flights or combining them with other probes could lead to meaningful upper bounds.

As our current AQN calculations focus on high-energy photon processes, future work could simulate the full low-energy photon propagation and reprocessing using a radiative transfer code such as \texttt{CosmoTherm}. This would enable more accurate predictions in the observational domain of BISOU and allow us to build matched-filter techniques or principal component analyzes tuned to AQN templates. Such work would enhance the scientific return of balloon-borne missions in constraining or discovering exotic dark matter scenarios like the AQN model.

\acknowledgments

FM, XL, MS, LVW, and AZ acknowledge the support from NSERC.
We thank Nils Schöneberg, Julien Lesgourgues, and Sven Guenther for helpful discussions and practical advice regarding the \texttt{CLASS} code, including troubleshooting installation issues and working with legacy versions. We are also grateful to the broader \texttt{CLASS} and \texttt{Cobaya} developer and user community for their publicly available documentation and tools.

\newpage
\appendix

\section{A check of screening effect}
\label{app:A check of screening effect}
When dealing with the electric field in plasma, it is important to also verify the presence of the Debye screening effect. The screening suppresses ${\cal Q}_{\rm eff}$ by an exponent $e^{-R_{\rm eff}/\lambda_{\rm D}}$, where $\lambda_{\rm D}$ is the Debye length:

\begin{equation}
\begin{aligned}
&
\lambda_{\rm D}
=\sqrt{\frac{k_BT_{\rm gas}}{4\pi\hbar c\alpha n_{\rm b}}}
=2.29{\rm\,cm}\left(\frac{10^4}{1+z}\right)\,,
\end{aligned}
\end{equation}
where we assume $T=2.73{\rm\,K}(1+z)$ and $n_{\rm b}=\rho_{\rm b}/m_p\approx2.482\times10^{-7}(1+z)^3{\rm\,cm^{-3}}$. Compared with the effective capture radius \eqref{eq:R_eff R 2} in plasma, we have: 
\begin{equation}
\begin{aligned}
\frac{R_{\rm eff}}{\lambda_{\rm D}}
&
=9.14\times10^{-4}
\left(\frac{1+z}{10^4}\right)^{5/4}
\left(\frac{m_{\rm AQN}}{100{\rm\,g}}\right)^{1/2}\qquad
{\rm (dense~plasma)}
\end{aligned}
\end{equation}
Therefore, the Debye screening is negligible in our present scenario ($10^3\lesssim z\lesssim10^6$). But this effect becomes considerable at $z\gtrsim10^8$ as suggested in  \cite{Flambaum:2018ohm,Ge:2019voa}.

\section{Weighted cross section $\sigma$}
\label{app:Weighted cross section sigma}

\subsection{Formulating $\sigma$}
From the definition \eqref{eq:dot Q} of $\sigma$, the weighted cross section $\sigma$ must be sensitive to the ionization fraction $X_e$. This is because in a fully ionized environment ($X_e \gtrsim 1$), $\sigma$ traces the effective cross section, $\sigma \approx \pi R_{\rm eff}^2$, while in a fully neutral environment ($X_e = 0$), $\sigma$ shrinks to the size of the geometrical cross section, $\sigma \approx \pi R^2$. Additionally, $\sigma$ must be sensitive to $Y_p$ because the weighted cross section is different in a hydrogen-rich environment from that in a helium-rich environment.

To find $\sigma$ as a function of $X_e$ and $Y_p$, we first re-express its definition \eqref{eq:dot Q} as:
\begin{equation}
\frac{\sigma(X_e,Y_p)}{\pi R^2}
=\frac{1}{\dot{Q}_0}\sum_i\dot{Q}_i
=\sum_i f_i\left(\frac{\sigma_i}{\pi R^2}\right)
\left(\frac{E_{{\rm ann},i}}{E_{\rm ann}}\right)
\left(\frac{n_i}{n_{\rm b}}\right)
\left(\frac{{\rm\Delta v}_i}{\rm\Delta v}\right)\,.
\end{equation}
As presented in table \ref{tab:Modification factors for each species}, most of the parameters here are well-known:
\begin{itemize}
    \item $f_i$: the efficiency of $p\bar{p}$ annihilation -- as discussed in section \ref{subsec:AQN energy injection in the early Universe}, neutral atoms have an additional suppression factor of 0.1 compared to ions;
    \item $\sigma_i$: ionized cross sections can be several orders of magnitude larger than the neutral case [e.g. eq. \eqref{eq:R_eff R 2}]\,\footnote{\label{footnote:enhancement by charge Z}In addition, ionized cross section can be enhanced by a factor of $\sqrt{Z}$, where $Z$ is the charge of the species. To derive this, one can replace eq. \eqref{eq:alpha hbar c Q_eff R_eff} with:
    \begin{equation*}
    \frac{\alpha\hbar c ZQ_{\rm eff}}{R_{\rm eff}}
    \approx k_{\rm B} T\,.
    \end{equation*}
    and perform the same treatment in section \ref{subsec:Characteristics of AQNs}, which gives $R_{\rm eff}^2\propto\sqrt{Z}$ in dense plasma.};
    \item $E_{{\rm ann},i}$: the annihilation energy -- helium produces more energy than hydrogen does due to a heavier mass;
    \item ${\rm\Delta v}_i$: a heavier element moves slower by a factor of the square root of its mass.
\end{itemize}

The only unknown term is the number density $n_i$, which is determined by the relation:
\begin{equation}
\frac{n_{\rm H,tot}}{n_{\rm b}}=1-Y_p\,,\qquad
\frac{n_{\rm He,tot}}{n_{\rm b}}=\frac{1}{4}Y_p\,,
\end{equation}
where $n_{\rm H,tot}$ and $n_{\rm He,tot}$ correspond to the total number densities of all states (neutral and ionized) of hydrogen and helium-4, respectively. The remaining problem is that each specific $n_i/n_{\rm b}$ still has complicated dependencies on $X_e$ and $Y_p$.

\begin{table}[h]
  \centering
  \caption{Parameters in $\dot{Q}_i/\dot{Q}_0$}
  \begin{tabular}{lccccc}
        \hline\hline
        Species & $f_i$ & $E_{{\rm ann},i}/E_{\rm ann}$ &  $\sigma_i/\sigma_0$  & $v_i/v$  \\
        \hline
        ~\,H    &     0.1     &   1            & 1         &  1        \\ 
        ~\,$\rm H^+$    &    1      &   1        &  $\left(\frac{R_{\rm eff}}{R}\right)^2$        &   1        \\ 
        $\rm ^4He$    &     0.1     &   4         &  1        &  $\frac{1}{2}$         \\ 
        $\rm ^4He^{+}$    &   1       &    4        & $\left(\frac{R_{\rm eff}}{R}\right)^2$         &  $\frac{1}{2}$        \\ 
        $\rm ^4He^{++}$    &  1        &   4         &  $\sqrt{2}\left(\frac{R_{\rm eff}}{R}\right)^2$        &  $\frac{1}{2}$ \\\hline\hline
  \end{tabular}
\label{tab:Modification factors for each species}
\end{table}

As suggested in cosmological simulations, such as RECFAST \cite{Seager:1999bc}, we can apply a simple approximation of $n_i$ in three cosmological epochs where $X_e$ remains relatively stable (i.e., all helium atoms settle in one specific ionization state). Specifically, it corresponds to $X_e=1$ ($\rm H^+$ and $\rm ^4He$), 1.08 ($\rm H^+$ and $\rm ^4He^+$), and $1.16$ ($\rm H^+$ and $\rm ^4He^{++}$). In the latter two epochs, $X_e$ can be expressed in terms of $Y_p$:
\begin{subequations}
\begin{equation}
X_e
=1+\frac{1}{4}\frac{Y_p}{1-Y_p}
\approx1.08\quad\implies\quad
n_{\rm H,tot}\approx n_{\rm H^+}\,,~
n_{\rm He,tot}\approx n_{\rm ^4He^+}\,;
\end{equation}
\begin{equation}
X_e
=1+\frac{1}{2}\frac{Y_p}{1-Y_p}
\approx1.16\quad\implies\quad
n_{\rm H,tot}\approx n_{\rm H^+}\,,~
n_{\rm He,tot}\approx n_{\rm ^4He^{++}}\,,
\end{equation}
\end{subequations}
where we choose $Y_p\approx0.245$. In the case of $X_e\leq1$, the expression can be even simpler:
\begin{equation}
X_e
\leq1
\quad\implies\quad
X_e n_{\rm H,tot}\approx n_{\rm H^+}\,,~
n_{\rm He,tot}\approx n_{\rm ^4He}\,.
\end{equation}
With these approximations, we can write a general form of a weighted cross-section that separates the ionized cross-section and the neutral cross-section:
\begin{subequations}
\label{eqs:sigma X_e Y_p pi R 2 etc.}
\begin{equation}
\frac{\sigma(X_e,Y_p)}{\pi R^2}
= F(X_e,Y_p)\left(\frac{R_{\rm eff}}{R}\right)^2+G(X_e,Y_p)\,, 
\end{equation}
\begin{equation}
\label{eqs:F X_e Y_p etc.}
F(X_e,Y_p)
\approx(1-Y_p) X_e\,,\quad
G(X_e,Y_p)
\approx0.1\left[(1-Y_p)(1-X_e)+\frac{1}{2}Y_p\right]\,,
\end{equation}
\end{subequations}
where $F(X_e,Y_p)$ and $G(X_e,Y_p)$ correspond to the (unknown) cross-section modification from ionized and neutral gases, respectively. The approximation \eqref{eqs:F X_e Y_p etc.} is almost exact when $X_e\leq1$. The approximation is still close to the exact values for larger $X_e$ as presented in table \ref{tab:Numerical values of F(X_e) and G(X_e)}. We conclude eq. \eqref{eqs:sigma X_e Y_p pi R 2 etc.} is a good approximation in the early Universe studied in this work.

\begin{table}[h]
  \centering
  \caption{Exact and approximate numerical values of $F(X_e,Y_p)$ and $G(X_e,Y_p)$ in three specific cosmological epochs ($Y_p=0.245$).}
  \begin{tabular}{c c c c c c} 
   \hline\hline
   $X_e$ & Species & $F(X_e,Y_p)$ & $F$-Approx & $G(X_e,Y_p)$ & $G$-Approx  \\ [0.5ex] 
   \hline
   1 & $\rm H^+$, $\rm ^4He$ & 0.755 & 0.755 & 0.0123 & 0.0123 \\ 
   1.08 & $\rm H^+$, $\rm ^4He^+$ & 0.878 & 0.815 & 0 & 0.0062  \\
   1.16 & $\rm H^+$, $\rm ^4He^{++}$ & 0.928 & 0.876 & 0 & 0.0002 \\
   \hline\hline
  \end{tabular}
  \label{tab:Numerical values of F(X_e) and G(X_e)}
\end{table}

\subsection{Enhancement of $\sigma$ from recursion}
From the previous subsection, we formulate $\sigma$ assuming that the background $X_e$ is insensitive to the energy injection from the AQNs. However, one may conjecture that a potential enhancement from a recursive loop: the background $X_e$ is enhanced by the energy injection from the AQNs, so that the weighted cross section is further amplified due to $\sigma\propto X_e$; a larger $\sigma$ results in more annihilation and energy injection into the background $X_e$. The positive feedback loop may largely enhance the estimation \eqref{eq:sigma pi R 2} of $\sigma$ and even trigger a runaway explosion of $X_e$. 

In this subsection, we verify that such an enhancement from recursion is negligible and cannot modify the estimation \eqref{eq:sigma pi R 2} of $\sigma$. Without loss of generality, we assume the Universe consists of only hydrogen and no heavier atoms such as helium. The recursive loop, if it ever exists, can only occur after recombination, where the background $X_e$ transits from unity to almost zero. The dominant reaction during this epoch is $e^-+p\leftrightarrow {\rm H}+\gamma$. The dynamic process can be derived by the Boltzmann equation (see e.g., textbook \cite{dodelson2003modern}):
\begin{equation}
\label{eq:a-3 dn_ea3 dt}
a^{-3}\frac{\rmd(n_e a^3)}{\rmd t}
=n_e^{(0)} n_p^{(0)}
\alpha^{(2)}
\left(\frac{n_{\rm H}}{n_{\rm H}^{(0)}}-\frac{n_en_p}{n_e^{(0)}n_p^{(0)}}\right)
+\frac{\dot{Q}}{E_\textrm{x-ray}}
\end{equation}
where the last term describes a new source term due to the energy injection from the AQNs\footnote{Here we make an optimistic assumption that one x-ray photon ionizes one neutral hydrogen atom. Also, we neglect the dependence of $a$ in the energy injection term, which is a second-order correction.}, $a$ is the scale factor, $\alpha^{(2)}$ is the average recombination rate $[{\rm cm^3\,s^{-1}}]$, $E_\textrm{x-ray}\sim(1-10){\rm\,keV}$ is the typical energy of the AQN-emitted photon, $n_i$ and $n_i^{(0)}$ are the number densities of species defined as:
\begin{equation}
n_i
=g_i e^{\mu_i/k_{\rm B}T}
\int\frac{\rmd^3p}{(2\pi)^3}
e^{-E_i/k_{\rm B}T}\,,\quad
n_i^{(0)}
=g_i
\int\frac{\rmd^3p}{(2\pi)^3}
e^{-E_i/k_{\rm B}T}\,,
\end{equation}
where $E_i\equiv\sqrt{(m_ic^2)^2+(pc)^2}$ is the relativistic energy, $\mu_i$ is the chemical potential in the reaction, $m_i
$ is the particle rest mass, and $i$ runs over the species $e$, $p$, and H. In a Universe of only hydrogen atoms, we have $n_e=n_p=X_e n_{\rm b}$ and $n_{\rm H}=(1-X_e)n_{\rm b}$. The Boltzmann eq. \eqref{eq:a-3 dn_ea3 dt} can be simplified to 
\begin{equation}
\label{eq:dX_e dt 3H X_e}
\frac{\rmd X_e}{\rmd t}
+3H X_e
=(1-X_e)(\beta+\gamma)-n_{\rm b}\alpha^{(2)}X_e(X_e-\delta)
\end{equation}
where $H=\frac{\dot{a}}{a}$ is the Hubble function, $\beta$ describes the ionization rate $[\rm s^{-1}]$ in the conventional cosmological reaction (see e.g. textbook \cite{dodelson2003modern}), $\gamma$ and $\delta$ describes the additional ionization from the AQN heat injection:
\begin{subequations}
\begin{equation}
\alpha^{(2)}
=9.78\frac{\alpha^2\hbar^2}{m_e^2c}
\left(\frac{\epsilon_0}{k_{\rm B}T}\right)^{1/2}
\ln\frac{\epsilon_0}{k_{\rm B}T}\,,\quad
\beta
=\frac{\alpha^{(2)}}{\hbar^3}
\left(\frac{k_{\rm B}m_eT}{2\pi}\right)^{3/2}e^{-\epsilon_0/k_{\rm B}T}\,,
\end{equation}
\begin{equation}
\gamma
=0.1\pi R^2\frac{E_{\rm ann}}{E_\textrm{x-ray}}gn_{\rm AQN}
{\rm\Delta v}\,,\quad
\delta
=\frac{\pi R^2}{\alpha^{(2)}}
\frac{E_{\rm ann}}{E_\textrm{x-ray}}
\frac{n_{\rm AQN}}{n_{\rm b}}\left(\frac{R_{\rm eff}}{R}\right)^2
g{\rm\Delta v}\,,
\end{equation}
\end{subequations}
where $E_{\rm ann}=2{\rm\,GeV}$ is the energy of $p\bar{p}$ annihilation, and $\epsilon_0=13.6{\rm\,eV}$ is the binding energy of $ep$ combination. Using the relation $n_{\rm AQN}\propto(1+z)^3m_{\rm AQN}^{-1}$, $R\propto m_{\rm AQN}^{1/3}$, and the baryon number density $n_{\rm b}=2.48\times10^{7}{\rm\,cm^{-3}}$, we obtain:
\begin{subequations}
\begin{equation}
\begin{aligned}
\frac{\gamma}{3H}
&
\approx5.67\times10^{-9}\left(\frac{1+z}{600}\right)^{3/2}
\left(\frac{1{\rm\,keV}}{E_\textrm{x-ray}}\right)
\left(\frac{{\rm\Delta v}}{15{\rm\,km\,s^{-1}}}\right)
\left(\frac{100{\rm\,g}}{m_{\rm AQN}}\right)^{1/3}\,,
\end{aligned}
\end{equation}
\begin{equation}
\delta
\approx3.54\times10^{-8}\left(\frac{1+z}{600}\right)
\left(\frac{1{\rm\,keV}}{E_\textrm{x-ray}}\right)
\left(\frac{{\rm\Delta v}}{15{\rm\,km\,s^{-1}}}\right)\,,
\end{equation}
\end{subequations}
where we assume $H(z)\propto(1+z)^{3/2}$ in a matter-dominated era, and an additional logarithmic term of $\ln(1+z)$ in $\delta$ is neglected due to its weak sensitivity. 

The estimate indicates that both $\gamma$ and $\delta$ are negligible terms in eq. \eqref{eq:dX_e dt 3H X_e}. In conventional recombination, the ionization fraction has a magnitude of $X_e\gtrsim10^{-4}$. Both energy injection terms ($\gamma/3H$ and $\delta$) are 4 orders of magnitude smaller than the conventional $X_e$ terms. We conclude that the enhancement of $\sigma$ from recursion is no more than a percentage correction of ${\cal O}(10^{-4})$.

\bibliographystyle{JHEP}
\bibliography{JCAP_ref} 

\end{document}